\documentclass[a4paper,english,british,aps,prb,twocolumn,latin9,utf8,floatfix]{revtex4}
\usepackage[T1]{fontenc}
\usepackage[utf8]{inputenc}
\usepackage{amstext}
\usepackage{amssymb}
\usepackage{graphicx}

\makeatletter

\newcommand{\lyxdot}{.}

\@ifundefined{textcolor}{}
{%
 \definecolor{BLACK}{gray}{0}
 \definecolor{WHITE}{gray}{1}
 \definecolor{RED}{rgb}{1,0,0}
 \definecolor{GREEN}{rgb}{0,1,0}
 \definecolor{BLUE}{rgb}{0,0,1}
 \definecolor{CYAN}{cmyk}{1,0,0,0}
 \definecolor{MAGENTA}{cmyk}{0,1,0,0}
 \definecolor{YELLOW}{cmyk}{0,0,1,0}
}

\usepackage{natbib}
\usepackage{doi}
\usepackage{hyperref}

\makeatother

\usepackage{babel}
\begin{document}

\preprint{This line only printed with preprint option}

\title{Micromagnetic modelling of anisotropic damping in ferromagnet}

\author{Mykola Dvornik}

\email{Mykola.Dvornik@ugent.be}

\homepage{http://dynamat.ugent.be}

\affiliation{DyNaMat Lab, Ghent University, Krijgslaan 281/S1, 9000 Ghent, Belgium}

\author{Arne Vansteenkiste}

\affiliation{DyNaMat Lab, Ghent University, Krijgslaan 281/S1, 9000 Ghent, Belgium}

\author{Bartel Van Waeyenberge}

\affiliation{DyNaMat Lab, Ghent University, Krijgslaan 281/S1, 9000 Ghent, Belgium}
\begin{abstract}
We report a numerical implementation of the Landau-Lifshitz-Baryakhtar
theory, which dictates that the micromagnetic relaxation term obeys
the symmetry of the magnetic crystal, i. e. replacing the single intrinsic
damping constant with a tensor of corresponding symmetry. The effect
of anisotropic relaxation is studied in thin saturated ferromagnetic
disk and ellipse with and without uniaxial magneto-crystalline anisotropy.
We investigate the angular dependency of the linewidth of magnonic
resonances with respect to the given structure of the relaxation tensor.
The simulations suggest that the anisotropy of the magnonic linewidth
is determined by only two factors: the projection of the relaxation
tensor onto the plane of precession and the ellipticity of the later. 
\end{abstract}
\maketitle
Landau and Lifschitz\cite{Landau1935} and later Gilbert\cite{Gilbert1955,Gilbert2004}
introduced a phenomenological relaxation term in the equation of motion
of magnetic moments in ferromagnetic media. They suggested that the
magnetic losses are characterized by a single intrinsic damping constant
of relativistic nature. Both the Landau-Lifshitz and the Gilbert phenomenological
damping terms are essentially equivalent for low magnetic losses,
while the Gilbert damping term works better for large values of the
damping constant, as was pointed out by Kikuchi\cite{Kikuchi1956}.
These terms are now widely used for the description of magnetic relaxations
in magnetic thin films\cite{Yu2012,Joyeux2011} and patterned magnetic
media\cite{Kruglyak2010}. The microscopic mechanism behind the magnetic
losses has also been suggested, e.g. in Gilmore et al.\cite{Gilmore2007}.

However, recent experimental data urges the development of new approaches
to magnetic losses, i.e. by introducing higher order terms within
Gilbert approach\cite{Tiberkevich2007}, by introducing inert relaxation\cite{Faehnle2011}
and by generalizing the magnetization dynamics and relaxation within
the framework of Onsager's kinetic equations\cite{Baryakhtar1998}.
The latter approach shows that the relaxation part of the equation
of precession should obey the crystallographic symmetry of the media.
Thereby replacing the single intrinsic damping constant with a tensor.
The reason behind anisotropic relaxation in magnetic media is the
symmetry properties of the spin-orbit and s-d interaction\cite{Heine1967},
which couple spins to other subsystems (i.e degrees of freedom), i.e.
lattice and free electrons, respectively. The anisotropic character
of these couplings results in anisotropic energy scattering from spins
to other subsystems and vice-versa. It worth noting, that an angular
dependency of magnetic losses (or effects associated with the same
physics) have already been reported experimentally\cite{Keatley2011,Serrano-Guisan2008,Schumacher2007,Mizukami2000,Schneider2007,Buchmeier2006}.
However analytical and numerical approaches still have to be developed,
especially for nano-scale structures where magnetic resonances are
strongly confined, and so, their spectra are discrete. 

In this paper, we report on a numerical implementation of Baryakhtar
theory within the mumax2 micromagnetic framework\cite{Vansteenkiste2011}.
Furthermore, we systematically investigate the influence of the anisotropic
relaxation on the the angular dependency of FMR linewidths in a nanoscale
magnetic disk and ellipse.

We start from the general Baryakhtar equation (LLBr)
\begin{equation}
\frac{\mathrm{\partial\mathbf{M}}}{\mathbf{\mathrm{\partial}\mathrm{t}}}=-\gamma\mathbf{M}\times\mathbf{H}+\lambda_{ij}(M)H_{j}+\lambda_{ij,sp}^{(e)}(M)\frac{\partial^{2}H_{j}}{\partial x_{s}\partial x_{p}}\label{eq:1}
\end{equation}
where $\mathbf{M}$, $\mathbf{\gamma}$, $\mathbf{H}$ are the magnetization
vector, gyromagnetic ratio and effective internal field respectively
(having contributions from exchange, spin-orbit, Zeeman and magneto-static
energies). $\mathbf{\lambda}$ and $\mathbf{\lambda^{\mathrm{(e)}}}$
are the relaxation tensors of relativistic and exchange nature, respectively,
and in general are functions of the magnetization. It is worth noting,
that in contrast to the Landau-Lifshitz formalism, the Baryakhtar
equation does not conserve the length of the magnetization vector,
i.e. $\left|\mathbf{M}\right|\neq const$. However in the present
study, we do not excite high-frequency (thermal) magnons and work
in linear approximation only, so that the vibrations of magnetization
vector length are negligible (although correctly accounted in our
micromagnetic simulations) . 

For the sake of simplicity, this paper focuses on the lowest magnetic
resonances (nearly uniform FMR) for which the second relaxation term
in eq. (\ref{eq:1}) (of exchange nature) is vanishing, i.e. $\mathbf{\lambda^{\mathrm{(e)}}\mathrm{\textrm{\ensuremath{\Delta}}}}\mathbf{H\ll\lambda H}$
. Such an assumption is reasonable since the majority of modern state-of-the-art
experimental techniques have limited (if any) spatial resolution \cite{Keatley2011,Perzlmaier2005},
and so, can only measure lowest magnonic resonances. In this case
eq. (\ref{eq:1}) reduces to 
\begin{equation}
\frac{\mathrm{\partial\mathbf{M}}}{\mathbf{\mathrm{\partial}\mathrm{t}}}=-\gamma\mathbf{M}\times\mathbf{H}+\lambda_{ij}(\mathbf{M})H_{j}\label{eq:2}
\end{equation}
The exact form of the relativistic tensor is unknown for an arbitrary
magnetic configuration. However, the tensor can be expanded into the
Taylor series in the vicinity of paramagnetic state ($\mathbf{M}=0)$\cite{Baryakhtar2010},
i. e. $\lambda_{ij}(\mathbf{M})=\lambda_{ij}(0)+\mu_{ij,sp}(0)M_{s}M_{p}+...$,
where $\lambda_{ij}$ and $\mu_{sp}$ are first and second order relativistic
relaxation tensors and $s,p$ denote spatial components which are
different from $i,j$. According to Ref. \cite{Baryakhtar2010}, the
first term in the expansion describe non-conservative relaxations,
where energy dissipations is accompanied with angular momentum transfer.
This mechanism dominates in ferromagnetic metals as become evident
from experiments on ultra-fast demagnetisation\cite{Koopmans2010},
therefore we omit higher order terms in Taylor expansion, so that
\foreignlanguage{english}{\textbf{$\lambda_{ij}(\mathbf{M})\mathbf{}=\lambda_{ij}(0)\mathbf{}$}}.
The latter expression could be rewritten in terms of dimensionless
damping tensor $\alpha_{ij}$, i.e. \foreignlanguage{english}{\textbf{$\lambda_{ij}=\gamma\left|\mbox{\textbf{M}}\right|\mathbf{M}\times\alpha_{ij}$}}.
So, finally eq. (\ref{eq:2}) transforms into 
\begin{equation}
\frac{\mathrm{\partial\mathbf{M}}}{\mathbf{\mathrm{\partial}\mathrm{t}}}=-\gamma\mathbf{M}\times\mathbf{H}+\gamma M\alpha_{\mathit{ij}}\mathbf{H}\label{eq:3}
\end{equation}
which under assumption of constant magnetisation length and isotropic
relaxations transforms into well-known Landau-Lifshitz equation, $\frac{\mathrm{\partial\mathbf{M}}}{\mathbf{\mathrm{\partial}\mathrm{t}}}=-\gamma\mathbf{M}\times\mathbf{H}+\gamma M\alpha\mathbf{M\times M}\times\mathbf{H}$.
The eq. (\ref{eq:3}) is employed in the present study.

The simulations were carried on a thin magnetic disk and ellipse with
equal thicknesses of 7 nm. The diameter of the disk is 110 nm, while
ellipse has minor and major axes of 44 nm and 110 nm, respectively.
The magnetic parameters are close to that of Cobalt, i.e. a saturation
magnetization of $\left|\mathbf{M}\right|=M_{s}=1440\cdot10^{3}\mathrm{\; A/m}$,
exchange stiffness constant of $A_{ex}=2.1\cdot10^{-11}\;\mathrm{J/m}$
and uniaxial anisotropy strength of $K{}_{u}=5.2\cdot10^{5}\;\mathrm{J/m^{3}}$(varying
in some of the simulations). Cobalt has a hexagonal lattice for which
the relaxation tensor is diagonal $\alpha_{ij}=\alpha\nu_{ij}\delta_{ij}$\cite{Baryakhtar1998,Baryakhtar2010}.
For the sake of simplicity we show its diagonal elements as a vector.
Cobalt is characterized by a relatively strong spin-orbit coupling,
making it an ideal candidate for our case study and for possible experiments.
The damping constant $\alpha$ is fixed to 0.008, while components
of the tensor $\nu_{ij}$ are varied to mimic given magnetic configurations.
In the present study,we only focus on the saturated case to depict
the main features of anisotropic damping. For this purpose in all
simulations we saturate the sample in-plane by using sufficiently
large applied magnetic field of $1\mathrm{T}$.

The magnonic spectra are extracted by means of FFT from 40 ns time
traces of the net magnetisation, simulated by exposing the relaxed
magnetic states to a ``sinc'' excitation with an amplitude of 100
Oe and cut-off of 80 GHz. The bandwidth of the simulations is maintained
at 100 GHz to avoid FFT aliasing. The magnetization dynamics always
vanish within the time-frame of the simulation. So we would not expect
any artificial broadening of the magnonic resonances. Nevertheless,
the a windowing function is also applied before FFT to prevent spectral
leakage. 

The dominant peaks in the spectra (attributed to the magnonic resonances
of different spatial characters) are fitted to Lorentzian curves in
order to extract their amplitudes, frequencies $\omega$ and FWHM
$\triangle\omega$. The latter two parameters are used to estimate
the relative net relaxation rates given by the Landau-Lifshits-Baryakhtar
and Landau-Lifshits models as $\Gamma_{LLBr}=\triangle\omega_{LLBr}/\omega_{LLBr}$
and $\Gamma_{LLG}=\triangle\omega_{LLG}/\omega_{LLG}$ , respectively.
Finally, the ratio between two is calculated as $\eta=\Gamma_{LLBr}/\Gamma_{LLG}$
to estimate the difference in two micromagnetic models.

\begin{figure}
\includegraphics[width=8cm]{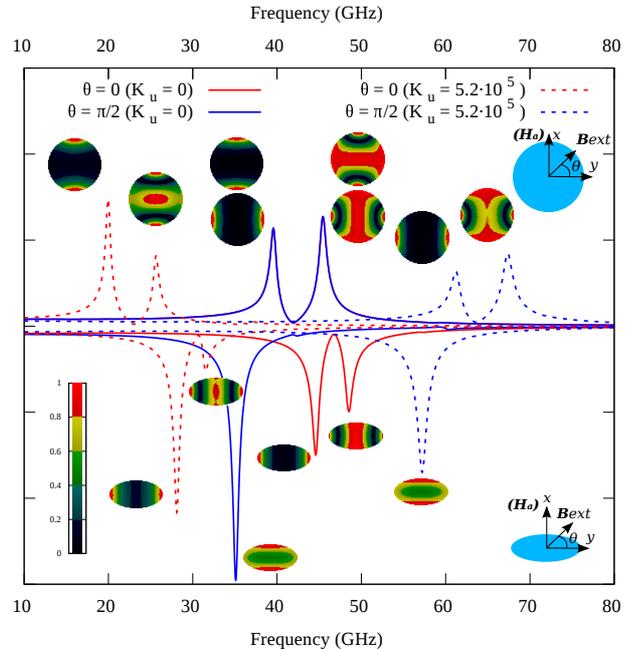}

\caption{The spectra of magnonic resonance and corresponding spatial mode profiles
are shown for disk (top panel) and ellipse (bottom panel). The solid
and dashed lines correspond to the isotropic and uniaxial materials,
respectively. \label{fig:example-spectra}}

\end{figure}

The spatial profiles of the magnonic modes are calculated using the
method from Ref. \cite{Dvornik2011a}. The typical spectra and spatial
profiles of the modes are shown in Fig. \ref{fig:example-spectra}.
For the disk, we excite two dominant modes of ``edge'' (lowest magnonic
mode) and ``bulk'' (higher order magnonic mode) character for any
direction of saturation (within $[0,\pi/2]$). In contrast, for an
ellipse only the edge mode remains within the frequency bandwidth
of simulations for all directions of saturation. The bulk mode is
rendered above 80 GHz due to enhanced contribution of exchange energy
when magnetised along minor axis. The uniaxial anisotropy acts as
expected from trivial physical considerations\cite{A.G.Gurevich1996},
i. e. hardening (softening) magnonic resonances for the parallel (perpendicular)
configurations, respectively.

\begin{figure}
\includegraphics[width=8cm]{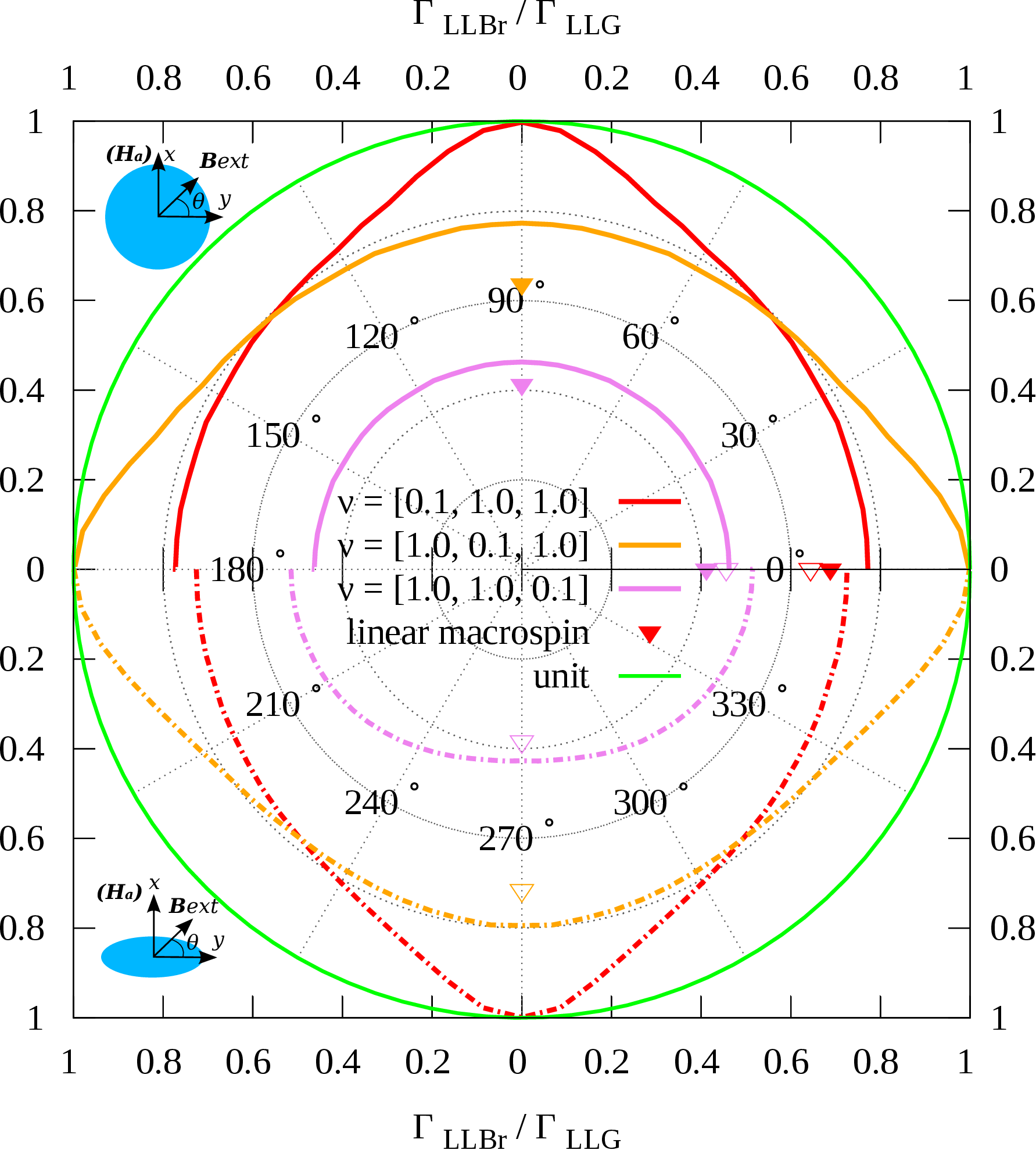}\caption{The angular dependency of the ratio between net relaxation rates of
LLBr and LLG is shown for different structures of damping tensor,
$\nu_{ii}$. The ratio is extracted for the edge mode in isotropic
Co nano-element ($K_{u}=0\: Jm^{-3})$. The top $(0,\pi)$ and bottom
$(\pi,2\pi)$ semi-planes correspond to the disk and ellipse, respectively.\label{fig:noaniz}}
\end{figure}

\begin{figure}
\includegraphics[width=8cm]{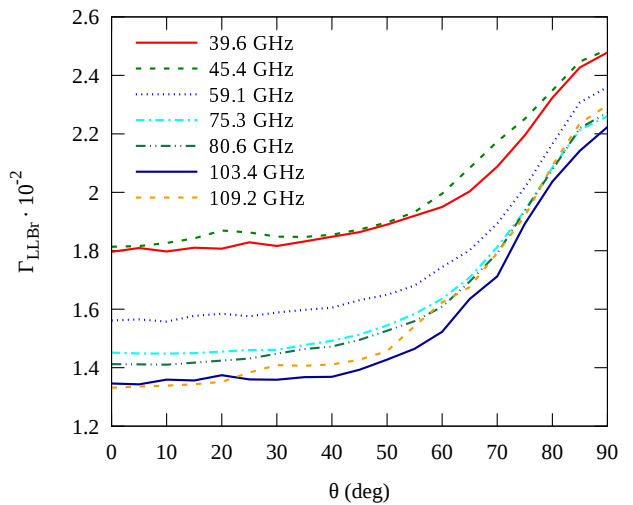}

\caption{The angular dependency of relative relaxation rate of different magnonic
modes in isotropic Co disk ($K_{u}=0\: Jm^{-3})$. is shown. The value
of the relaxation tensor is $\nu_{ii}=[0.1,1,1]$. The numbers in
the key represent frequencies of the modes expressed in GHz.\label{fig:modes} }
\end{figure}

The anisotropy of magnonic damping has never been measured qualitatively.
Since the aim of this paper is to show its effect on magnetisation
dynamics, the degree of anisotropy is chosen to make sure that simulations
are both computationally feasible and numerically stable. So in all
following cases, the anisotropy

of damping is represented by reducing given components of the relaxation
tensor by one order of magnitude. However, Baryakhtar showed that
components of the relaxation tensor could be expressed in terms of
magneto-crystalline anisotropy constants\cite{Baryakhtar2010}. So
the damping should be isotropic, when there is no magneto-crystalline
anisotropy present. For demonstration purposes we vary it nevertheless
in this study. Furthermore, in the simulations where anisotropy constant
is varied, the corresponding components of the damping tensor remain
constant to draw straight comparison between LLG and LLBr models.
More rigorous treatment is expected to change our results quantitatively,
but not qualitatively. 

The angular dependency of the relative macroscopic damping constant
for different structures of the tensor $\nu_{ii}$ is shown in Fig.
\ref{fig:noaniz} for an isotropic Co disk (top panel) and ellipse
(bottom panel). First of all, the symmetry of the graph is two-fold
and mimics that of the relaxation tensor. Secondly, the data suggests
that changes of the relaxation constant along saturation direction
do not change the value of the macroscopic damping constant with respect
to the one obtained from LLG, i.e. $\eta_{0}^{1,0.1,1}=\eta_{\pi/2}^{0.1,1,1}\approx1$,
where top and bottom indices show diagonal elements of the relaxation
tensor and angle of saturation respectively. Meanwhile, when the components
of the relaxation tensor are reduced orthogonally to the saturation
direction, a distinct reduction of relative relaxation rate is observed,
i.e. $\eta_{0}^{0.1,1,1}=\eta_{\pi/2}^{1,0.1,1}\thickapprox0.77$
and $\eta_{0}^{0.1,1,1}\thickapprox0.73,\:\eta_{\pi/2}^{1,0.1,1}\thickapprox0.79$,
for the disk and ellipse, respectively. Finally, for the case of reduced
out-of-plane component of relaxation tensor, $\nu_{ii}=(1,1,0.1)$,
the relative macroscopic damping is always reduced below the value
observed for the in-plane case, i.e. $\eta_{0}^{1,1,0.1}=\eta_{\pi/2}^{1,1,0.1}\thickapprox0.46$
and $\eta_{0}^{1,1,0.1}\thickapprox0.51,\;\eta_{\pi/2}^{1,1,0.1}\thickapprox0.43$,
for disk and ellipse respectively. So when the relaxation tensor is
uniform in-plane, the relative relaxation rate in the disk is independent
on the direction of saturation as it follows from its rotational symmetry.
In contrast, for the ellipse the symmetry is two-fold, so not only
the relaxation tensor contributes to the symmetry of the magnonic
damping.

In general, the magnetisation vector has three degrees of freedom.
Two of them correspond to precession (transverse degrees of freedom)
and one corresponds to the changes of the vector's length (longitudinal
degree of freedom). In linear approximation, the longitudinal degree
of freedom is ``frozen'', so that it does not dissipate energy.
It explains why there are no changes observed between LLBr and LLG,
when the relaxation tensor is altered along the longitudinal direction.
Linearisation of the LLBr equation in the macrospin approximation
under assumption of diagonal relaxation tensor leads to

\begin{equation}
\Gamma\thicksim\mathbf{\nu}_{ii}\frac{\omega_{\mathit{i}}}{\omega}=\mathbf{\mathbf{\nu}_{\mathit{ii}}\sqrt{\frac{\omega_{\mathit{i}}}{\omega_{\mathit{j}}}}}\label{eq:4}
\end{equation}
The frequency of the precession is given by $\omega^{2}=\omega_{i}\omega_{j},i\neq j$,
$\omega_{i}$ and $\omega_{j}$ are characteristic frequencies of
transversal degrees of freedom. Eq.(\ref{eq:4}) has clear physical
meaning, so that the amount of dissipated energy is proportional to
the energy stored in all degrees of freedom. So the striking difference
between relaxation rates, i. e. $\eta_{0}^{0.1,1,1}/\eta_{\pi/2}^{1,1,0.1}=1.67$
and $\eta_{0}^{0.1,1,1}/\eta_{0}^{1,1,0.1}=1.59$ and \foreignlanguage{english}{$\eta_{\pi/2}^{1,0.1,1}/\eta_{\pi/2}^{1,1,0.1}=1.84$,}
suggest that for the given magnonic mode the out-of-plane degree of
freedom stores more energy, than in-plane. So by estimating the ratio
between the relaxation rates $\Gamma/\Gamma^{'}$ for the different
structures of the damping tensor $\nu$ and $\nu^{'}$, it is possible
to find the characteristic frequency of different degrees of freedom.
In particular, for the out-of-plane component it reads: 
\begin{equation}
\omega_{z}=\sqrt{\frac{\nu_{ii}-(\Gamma/\Gamma^{'})\nu_{ii}^{'}}{(\Gamma/\Gamma^{'})\nu_{zz}^{'}-\nu_{zz}}}\omega,\; i=x,y\label{eq:freq}
\end{equation}
If our assumption is valid, then for the disk the frequency should
be independent of the direction of in-plane saturation. For the lowest
magnonic mode, the simulated relaxation rates for $\nu_{ii}=[0.1,1,1]$
and $\nu_{ii}'=[1,1,0.1]$, lead to the values of $\omega_{z}$ of
$\approx54.26$ GHz and $\approx47.94$ GHz, at 0 rad and $\frac{\pi}{2}$
rad, respectively. For the higher order magnonic mode, the values
of $\omega_{z}$ at 0 rad and $\frac{\pi}{2}$ rad are $\approx62.54$
GHz and $\approx55.54$ GHz, respectively. 

The relative difference of $\frac{\omega_{z}(0)-\omega_{z}(\pi/2)}{\omega_{z}(0)}\thickapprox11.2\%$
between the values of $\omega_{z}$ at 0 rad and $\frac{\pi}{2}$
rad is roughly the same for these two modes, suggesting presence of
damping mechanism beyond trivial linear macrospin model. The calculated
angular dependency cannot be simply attributed to the artificial edge
roughness (and corresponding two-magnon scattering relaxation mechanism),
because ``bulk'' modes are not sensitive to it\cite{Maranville2006}.
So we assume that the observed discrepancy is due to the excitation
of the second harmonics in the longitudinal degree of freedom. According
to the simulations, the ratios between the amplitudes of the second
harmonics and corresponding eigen modes are around $10^{-3}$. This
dissipation channel is not accounted in our simple model.

The analysis suggests that for the ``edge'' and ``bulk'' modes
of the disk, the out-of-plane characteristic frequencies are much
larger than the in-plane, $\omega_{z}\gg\omega_{y}$. The ratio between
the characteristic frequencies could be controlled by various types
of anisotropy, e.g. magneto-crystalline and shape anisotropies. Moreover,
the ratio between the $\omega_{z}$ and $\omega_{y}$ might change
for the higher-order modes because of interplay of shape anisotropy
and exchange energies as explained in Ref.\cite{Dvornik2011}. In
particular, in thin magnetic nano-elements $\omega_{y}/\omega_{z}\ll1$
and $\omega_{y}/\omega_{z}\rightarrow1$ for low- and high-frequency
magnons, respectively. Therefore we can expect that the relaxation
rate is mode specific. The relative relaxation rate, $\Gamma_{LLBr}$,
for different magnonic modes of the isotropic disk with $\nu_{ii}=[0.1,1,1]$
is presented in Fig. \ref{fig:modes}. The relative linewidth decreases
with the frequency of the mode as we expected. So for the higher order
magnons the in-plane characteristic frequency tends to out-of-plane,
thereby equalizing both relaxation channels. So for high frequency
magnons the relaxation rate eventually approaches the limit defined
by the components of the relaxation tensor, i.e. 0.0044 and 0.008
for $\theta=0$ and $\theta=\pi/2$, respectively. The calculated
values of the relaxation rate are somewhat higher than the intrinsic
values fixed by the relaxation tensor. This can be attributed to the
non-uniform spatial character of the magnonic modes\cite{Hillebrands2010}.

\begin{figure}
\includegraphics[width=8cm]{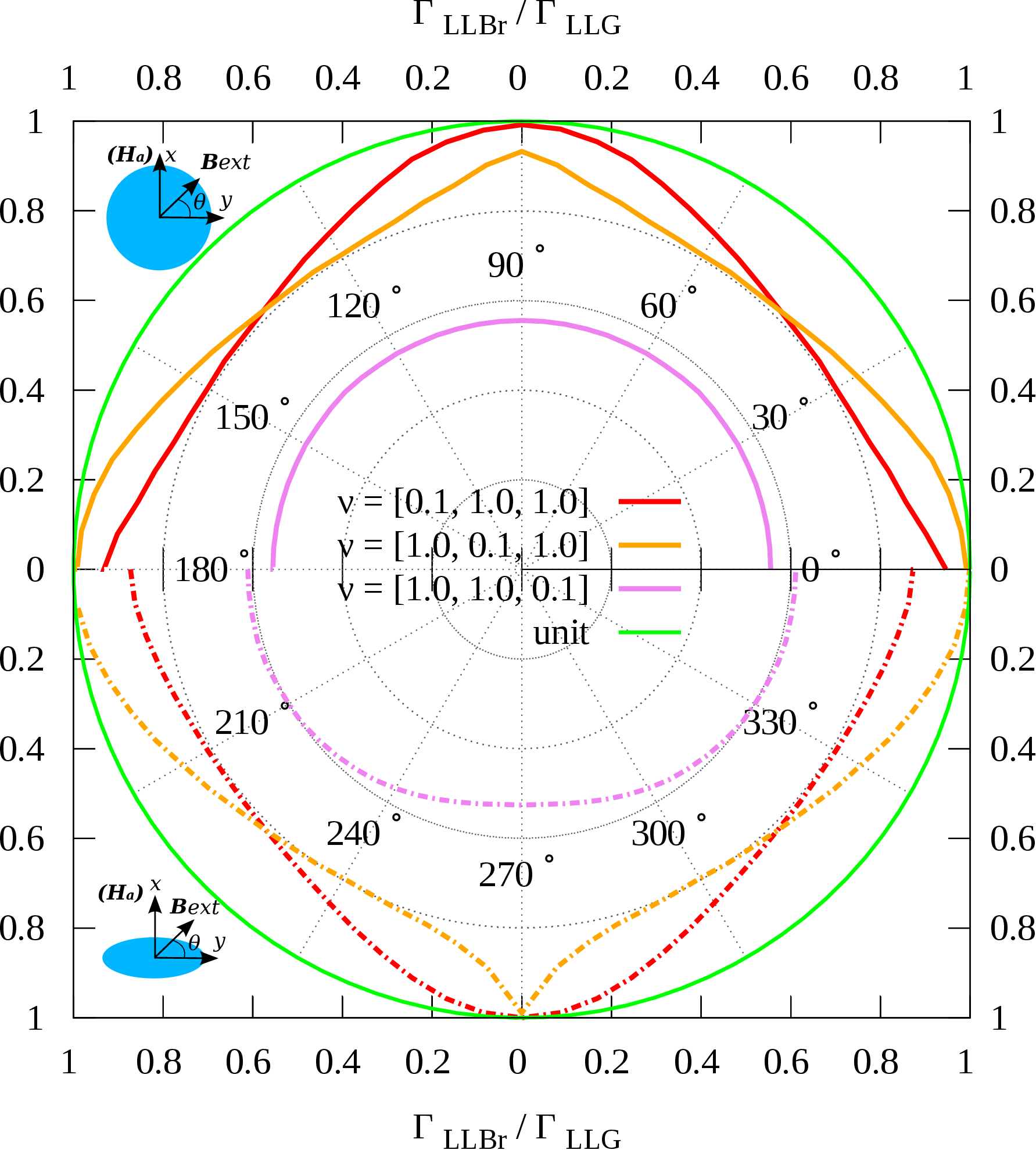}

\caption{The angular dependency of the ratio between net relaxation rates of
LLBr and LLG is shown for different structures of damping tensor,
$\nu_{ii}$. The ratio is extracted for the edge mode in uniaxial
Co nano-element ($K_{u}=5.2\cdot10^{5}\mathrm{\: Jm^{-3}})$. The
top $(0,\pi)$ and bottom $(\pi,2\pi)$ semi-planes correspond to
the disk and ellipse, respectively.\label{fig:aniz}}
\end{figure}

Lets us employ this formalism to explain the results obtained in an
ellipse, which naturally introduces an in-plane shape anisotropy.
For the cases of $\nu_{ii}=[0.1,1,1]$ at 0 rad (along minor axis)
and $\nu_{ii}=[1,0.1,1]$ at $\frac{\pi}{2}$ rad (along major axis)
the in-plane and out-of-plane characteristic frequencies are $\omega_{x}\approx36.14$
GHz, $\omega_{z}\approx55.11$ GHz and $\omega_{y}\approx23.81$ GHz,
$\omega_{z}\approx51.11$ GHz, respectively. For both cases the $\omega_{z}$
is similar to that of the disk, while the in-plane characteristic
frequency is enhanced (reduced) for saturation along major (minor)
axis. The latter is expected for in-plane shape anisotropy\cite{Pardavi-Horvath2011}.
For the quantity represented in Fig. \ref{fig:noaniz} we can simply
write:

\begin{equation}
\eta=\frac{\Gamma_{LLBr}}{\Gamma_{LLG}}=\frac{1}{1+\omega_{z}/\omega_{i}}\nu_{ii}+\frac{1}{1+\omega_{i}/\omega_{z}}\nu_{zz},\label{eq:5}
\end{equation}
where $i=x,y$ for 0 rad and $\frac{\pi}{2}$ rad respectively. These
theoretical values are given by dots in Fig. \ref{fig:noaniz}. The
theory qualitatively mimics the behaviour observed in Fig. \ref{fig:noaniz},
further supporting our linear model of anisotropic magnonic relaxation. 

\begin{figure}
\includegraphics[width=8cm]{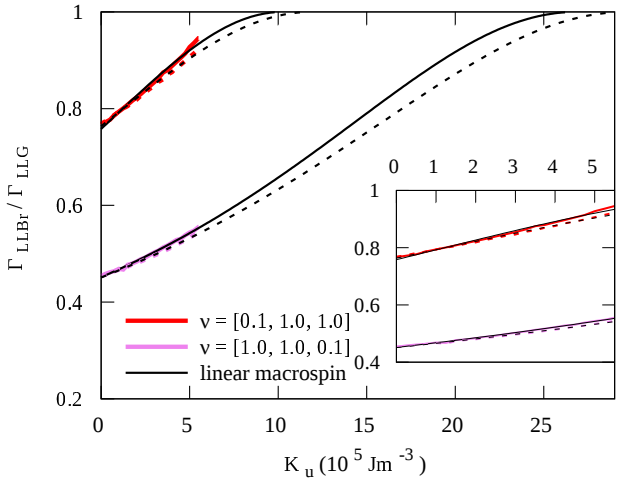}

\caption{The relative magnonic resonance linewidth $\Gamma_{LLBr}/\Gamma_{LLG}$
in function of uniaxial anisotropy constant $K_{u}$ is shown for
different structures of the relativistic relaxation tensor $\alpha\nu_{ii}$.
The solid and dotted lines correspond to the ``edge'' and ``bulk''
modes of uniaxial Co disk, respectively. As we increase the uniaxial
anisotropy constant, the relaxation rate calculated with LLBr tends
to that of LLG.\label{fig:Ku_dep}}
\end{figure}

In contrast to shape, magneto-crystalline anisotropy could be easily
introduced and tuned. The relative linewidth calculated on the same
samples, but in the presence of uniaxial anisotropy is shown in Fig.~\ref{fig:aniz}.
The direction of uniaxial anisotropy always coincides with the shortest
eigenvector of the damping tensor for the reasons explained above.
The results presented in Fig.~\ref{fig:aniz} qualitatively reproduce
those from Fig. \ref{fig:noaniz}. For all the cases at 0 ($\frac{\pi}{2}$)
rad the uniaxial anisotropy softens (harden) lowest magnonic mode.
Thus, reducing (enhancing) influence of in-plane relaxation. Therefore
the relative relaxation rate is always enhanced as compared to anisotropy-less
case. Furthermore, in case of ellipse for $\nu_{ii}=[1,0.1,1]$ at
$\frac{\pi}{2}$ rad the relative linewidth tends to unity. The effect
could be easily explained if we take into account that shape and uniaxial
anisotropies act in the same way, i.e. soften the lowest magnonic
mode, eventually vanishing the role of in-plane relaxation. Since
out-of-plane relaxation is equal to that of isotropic case, the linewidth
also tends to that of isotropic case.

So, by tuning the strength of the uniaxial anisotropy we can effectively
change the relaxation rate within the limits defined by the anisotropy
of damping tensor. The relative linewidth as a function of uniaxial
anisotropy constant, $K_{u}$, is shown in Fig. \ref{fig:Ku_dep}.
The fit to our simple model is also shown. We assumed that the frequency
of the mode is given by (SI units):

\begin{equation}
\omega=\omega_{0}-\frac{(\gamma/2\pi)}{\mu_{0}M_{s}}K_{u}
\end{equation}
where $\omega_{0}$ is the frequency of the mode in isotropic case
with the assumption that contribution of the anisotropy energy is
second order with respect to $\omega_{0}$. So the relative relaxation
rate is give by:

\begin{equation}
\eta=\frac{1}{1+(\omega_{j}/\omega)^{2}}\nu_{ii}+\frac{1}{1+(\omega/\omega_{j})^{2}}\nu_{jj}
\end{equation}
The relative linewidth increases with the strength of uniaxial anisotropy
and tends to unity, as it seen from our model. In particular, for
the given case of the the anisotropy axis parallel to either in-plane
or out-of-plane degrees of freedom, the mode frequency (and so corresponding
characteristic frequency) is reduced as compared to the isotropic
case. Thereby reducing the influence of relaxations along the anisotropy
axis (the one which is reduced as compared to LLG). Therefore, as
the the strength of the anisotropy increases, the relaxation rate
tends to that of the LLG. The significant difference between the slopes
in Fig. \ref{fig:Ku_dep} is due to the difference in characteristic
frequencies as it follows from eq. \ref{eq:5}.

Based on our model and results obtained in isotropic disk and ellipse
we conclude that (a) the highest reduction of relaxation rate is observed
when the shortest eigen vector of the relaxation tensor is parallel
to the hardest degree of freedom and (b) reduction of characteristic
frequency leads to the reduction of corresponding relative relaxation
rate. These conclusions could be used to design a representative experiment,
e.g. measurements of magnonic linewidths in Co nano-elements with
in-plane and out-of-plane uniaxial anisotropies. Then by estimating
the characteristic frequencies from the micromagnetic simulations,
the corresponding components of the relaxation tensor could be extracted.

We showed that anisotropic nature of the relaxation constant leads
to a corresponding anisotropy in the linewidth of magnonic resonances.
We showed that magnonic relaxation rate is mode specific even if the
relaxation tensor is uniform. The strength of the effect depends on
the ratio between the characteristic frequencies of transverse degrees
of freedom of magnetization. The introduction of anisotropic damping
makes this effect even more prominent. We showed that the relaxation
rate could be altered by changing the strength of the magneto-crystalline
anisotropy. We suggest that the latter effect could be experimentally
verified in multiferroic materials. Finally, our numerical implementation
is open, and so, it can be freely used by the community to fit experimental
data. 

\bibliographystyle{apsrev4-1}
\bibliography{/run/media/mykola/LaCie/PAPERS/references}

\end{document}